\documentclass{article}
\usepackage{spconf,amsmath,graphicx}
\usepackage[hidelinks]{hyperref} 
\usepackage{graphicx}
\usepackage{booktabs} 
\usepackage{array}
\usepackage{amssymb}
\usepackage{amsfonts}
\usepackage{caption}
\usepackage{subcaption} 
\usepackage[table,xcdraw]{xcolor}
\usepackage{colortbl}
\usepackage{enumitem}

\title{Transfer Learning for Paediatric Sleep Apnoea Detection Using Physiology-Guided Acoustic Models}
\name{
\begin{tabular}{c}
Chaoyue Niu$^1$, Veronica Rowe$^1$, Guy J. Brown$^1$, Heather Elphick$^2$, Heather Kenyon$^2$, \\
\textit{Lowri Thomas$^2$, Sam Johnson$^3$ and Ning Ma$^1$}
\end{tabular}
\thanks{The authors would like to thank PFL Healthcare (UK) Limited for providing the data collection infrastructure. The study was funded by UKRI MRC impact acceleration accounts (grant 182731) and NIHR Great Ormond Street Hospital Biomedical Research Centre (grant 187217).}
}
\address{
$^1$School of Computer Science, University of Sheffield, S1 4DP, UK.\\
$^2$Sheffield Children's NHS Foundation Trust, S10 2TH, UK.\\
$^3$Passion for Life Healthcare, Chester, CH1 2NP.\\
Email: \{c.niu, verowe1, g.j.brown, n.ma\}@sheffield.ac.uk, \\
\{h.elphick, lowrithomas, heather.kenyon2\}@nhs.net, sam.johnson@passionforlife.com.}

\begin{document}
\ninept
\maketitle
\begin{abstract}
Paediatric obstructive sleep apnoea (OSA) is clinically significant yet difficult to diagnose, as children poorly tolerate sensor-based polysomnography. Acoustic monitoring provides a non-invasive alternative for home-based OSA screening, but limited paediatric data hinders the development of robust deep learning approaches. This paper proposes a transfer learning framework that adapts acoustic models pretrained on adult sleep data to paediatric OSA detection, incorporating SpO\textsubscript{2}-based desaturation patterns to enhance model training. Using a large adult sleep dataset (157 nights) and a smaller paediatric dataset (15 nights), we systematically evaluate (i) single- versus multi-task learning, (ii) encoder freezing versus full fine-tuning, and (iii) the impact of delaying SpO\textsubscript{2} labels to better align them with the acoustics and capture physiologically meaningful features. Results show that fine-tuning with SpO\textsubscript{2} integration consistently improves paediatric OSA detection compared with baseline models without adaptation
%across the whole data set in terms of MAE and RMSE measures
. These findings demonstrate the feasibility of transfer learning for home-based OSA screening in children and offer its potential clinical value for early diagnosis.
\end{abstract}

\begin{keywords}
Sleep Disordered Breathing, Paediatric Obstructive Sleep Apnoea, Acoustic Model, Transfer Learning, Physiological Delay Modelling
\end{keywords}

\section{Introduction}
\label{sec:intro}

Obstructive sleep apnoea (OSA) is a common paediatric sleep-disordered breathing condition characterised by repeated uppper-airway obstruction, leading to intermittent hypoxia, disrupted ventilation, and fragmented sleep~\cite{kaditis2015obstructive}. Its prevalence is 1–5\% in the general paediatric population~\cite{marcus2012diagnosis}, rising to 33–61\% among children with obesity~\cite{roche2020obstructive}. Untreated OSA is associated with cardiovascular, metabolic, and neurocognitive impairments, behavioural problems, growth delay, and reduced quality of life~\cite{qin2024pediatric}.

Polysomnography (PSG) is the diagnostic gold standard, but requires overnight lab-based monitoring with specialised equipment and personnel, resulting in high costs and low tolerability in children. Although effective treatments exist~\cite{kheirandish2010abnormal, gozal2020treatment}, the complexity and cost of PSG limit its use as a universal diagnostic tool, particularly for children with habitual snoring who are at risk of OSA~\cite{tan2015pediatric}. Consequently, over 90\% of OSA cases in children remain undiagnosed~\cite{alexander2019rapid}, and accessible, affordable, minimally invasive diagnostics is urgently needed~\cite{roebuck2015comparison}.  

Alternative approaches, such as questionnaires~\cite{kadmon2013validation}, symptom-based scoring~\cite{chang2013combination}, and simplified recordings~\cite{lazaro2013pulse}, lack sufficient diagnostic accuracy~\cite{bertoni2019towards}. Machine learning (ML) shows promise for automated, objective paediatric OSA diagnosis~\cite{obermeyer2016predicting}, and acoustic monitoring offers a low-cost, non-invasive option for home-based OSA screening~\cite{romero2022acoustic, hussain2019cost}. 
However, paediatric applications remain limited due to weaker respiratory sounds, developmental physiology variability, and scarcity of labelled datasets.

Furthermore, OSA data collection is constrained by the slow and labour-intensive process of manual scoring. Paediatric data are particularly scarce, as children have lower tolerance for sensors, which further limits model development.
Transfer learning~\cite{vrbanvcivc2020transfer} provides a practical solution by adapting models pretrained on large adult datasets to the paediatric domain. While successfully applied to sleep stage prediction from adults to children \cite{haimov2024deep}, its use in acoustic-based OSA detection and integration of physiological information, such as oxygen desaturation, has not yet been explored.

This study proposes transfer learning for paediatric sleep apnoea detection by adapting a convolutional neural network (CNN) trained on adult acoustic data. The framework supports both a single-task head predicting OSA and a multi-task head predicting OSA and time below a night-specific oxygen saturation (SpO\textsubscript{2}) threshold. Two adaptation strategies are investigated: freezing the encoder versus full fine-tuning. Physiological priors are incorporated by modelling the temporal delay between nasal airflow cessation and oxygen desaturation. This work makes three main contributions:
(i) A transfer learning framework for acoustic-based paediatric OSA detection using adult-pretrained models, (ii) A systematic comparison of single- and multi-task learning under different fine-tuning strategies, and (iii) Integration of physiological priors via SpO\textsubscript{2} desaturation delay modelling to improve clinically meaningful severity estimation.

The rest of the paper is organised as follows. Section~\ref{s:data} describes the data used in this study. Section~\ref{s:methods} presents the proposed transfer learning framework for acoustic-based paediatric OSA detection and integration with SpO\textsubscript{2} information. Experiments are described in Section~\ref{s:expr}. Results and Discussion are given in Section~\ref{s:results}. Section~\ref{s:conc} concludes this paper and presents future directions.

\begin{figure*}[t]
    \centering
    \includegraphics[width=.98\linewidth]{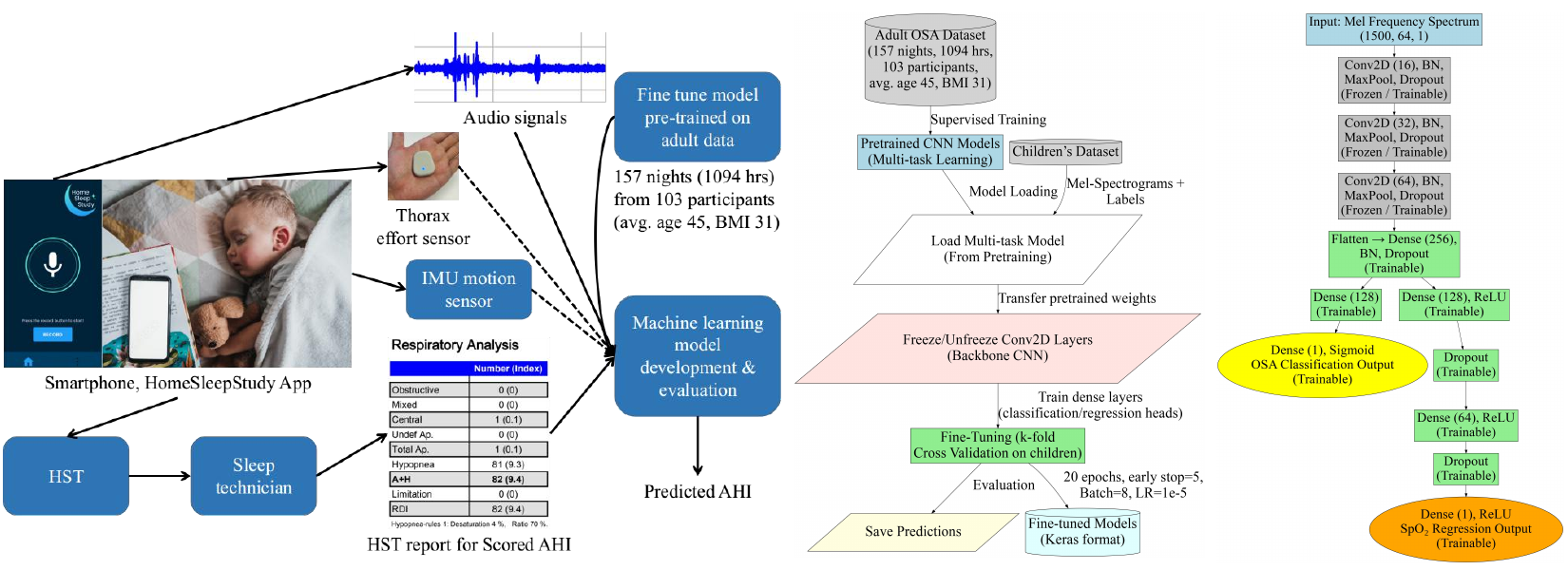}
    % \caption{Pipeline of data collection and model development: breathing sounds are recorded via smartphone, optionally with a chest sensor and IMU, and clinician-annotated HST provides apnoea–hypopnoea events and scored AHI. Solid lines show data used in the current model; dashed lines indicate unused data for future integration. The fine-tuning pipeline loads an adult-pretrained multi-task CNN, incorporates paediatric Mel-spectrograms and labels, optionally freezes the shared CNN backbone, and fine-tunes the task-specific dense and output layers. }
    \caption{Pipeline for data collection and model development. Breathing sounds are recorded using a smartphone, optionally supplemented with chest sensor and IMU signals. Clinician-annotated HST provides apnoea–hypopnoea events and AHI scores. Solid lines denote data used in the current model, while dashed lines indicate signals reserved for future integration. The fine-tuning pipeline loads an adult-pretrained multi-task CNN, optionally freezes the shared CNN backbone, and fine-tunes task-specific layers.}
    \label{fig:Pipeline}
\end{figure*}

%\section{Materials and Methods}
\section{Data}
\label{s:data}

Paediatric sleep data were collected by Sheffield Children’s NHS Foundation Trust with ethics approval from the UK NHS Research Ethics Committees (REC reference: 24/PR/0284). 15 children (age range 1--15 years, mean $6 \pm 4$; 5 females, 10 males) were recruited for overnight monitoring. Obesity, a key OSA risk factor, is assessed via body mass index (BMI), calculated as $\text{Weight (kg)}/\text{Height (m)}^2$. BMI was recorded for all participants (mean BMI $25 \pm 10$). Participant demographics are summarised in Table~\ref{Tab:15_results}.

Data were acquired using a Home Sleep Apnoea Test (HSAT) device (SOMNOtouch\texttrademark{} RESP\footnote{\url{https://somnomedics.de/en/solutions/sleep_diagnostics/polygraphy-devices/somnotouch-resp/}}), alongside concurrent smartphone audio recordings in both laboratory and home settings. The HSAT system recorded SpO\textsubscript{2}, heart rate, airflow, respiratory effort, snoring, sleep–wake status, and body position, with an optional thorax sensor. Audio was recorded at 16\,kHz/16-bit resolution in 2-minute segments via a custom mobile app (Figure~\ref{fig:Pipeline}). Apnoea events were manually scored by a registered polysomnographic technologist\footnote{\url{https://www.brpt.org/}}. In total, $\sim$120 hours of data were obtained across 15 nights. Audio quality was assessed by the snore-to-non-snore ratio, while HSAT signal quality was evaluated by the percentage of missing SpO\textsubscript{2}.  

The severity of OSA is clinically assessed using the apnoea-hypopnoea index (AHI), defined as the number of apnoea and hypopnoea events per hour of sleep.
%$\text{AHI} = (N_{\text{apnea}} + N_{\text{hypopnea}})/T_{\text{sleep}}$, where $T_{\text{sleep}}$ is sleep time in hours. 
In children, OSA severity is classified as normal/mild (AHI $<$ 5), moderate (AHI = 5--10), or severe (AHI $>$ 10)~\cite{gourishetti2021baseline}. 

% The adult dataset comprised 103 participants (67 males, 36 females; mean age 45 $\pm$ 13 years; mean BMI 31 $\pm$ 7), providing 157 nights and $\sim$1094 hours of recording (see \cite{romero2022acoustic} for more details).  
For comparison, the adult dataset comprised 103 participants (67 males, 36 females; mean age $45 \pm 13$ years; mean BMI $31 \pm 7$), contributing 157 overnight recordings and approximately 1,094 hours of data. Further details of the adult cohort and data collection protocol are reported in \cite{romero2022acoustic}.

\section{Methods}
\label{s:methods}

\subsection{Feature extraction and preprocessing}

\begin{figure}[thb]
    \centering
    \includegraphics[width=.95\linewidth]{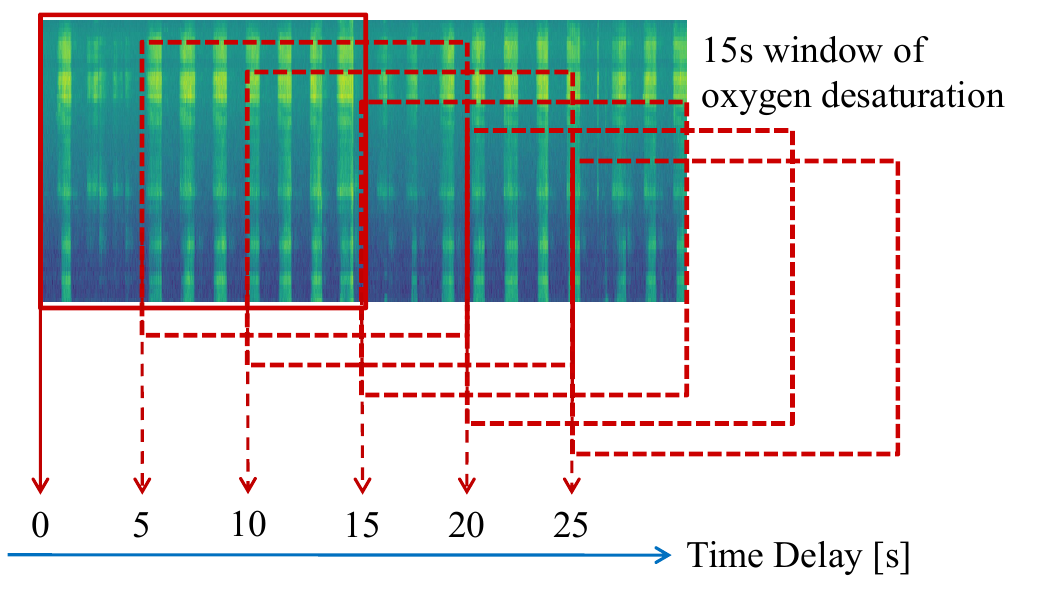}
        \caption{Time delay of oxygen desaturation and Mel-frequency spectrum. Red dashed rectangles mark the 15-s oxygen desaturation window, which is shifted from 0 to longer time delays to identify the delay that best discriminates between normal and abnormal breathing.}
        \label{fig:delay}
\end{figure}

Full-night audio recordings were segmented into overlapping 30\,s windows $\mathcal{S}_i$ with a 10\,s shift. Each segment is processed via short-term Fourier transform (50\,ms Hann window, 20\,ms hop) to obtain a power spectrogram $P(t,f)$. A 64-filter Mel-filterbank is applied to produce log-compressed Mel-spectrograms $M(t,m)$, followed by normalisation per Mel-bin to obtain $\hat{M}_i \in \mathbb{R}^{T \times F}$. These features serve as input to a CNN $f_\theta$ for binary OSA detection $y_i \in \{0,1\}$, with $y_i = 1$ segments grouped into events and aligned with HSAT annotations. For physiological integration, oxygen desaturation was quantified as the percentage of time below the night-specific SpO\textsubscript{2} baseline (3\% drop) according to AASM guideline\footnote{American Academy of Sleep Medicine guideline: \url{https://aasm.org/clinical-resources/practice-standards/practice-guidelines/}} within a 15\,s window for each 30\,s segment (see Fig.~\ref{fig:delay}).

\subsection{Problem formulation}
Let the adult dataset be
\begin{equation}
\mathcal{D}_A = \{(X_i^A, y_i^A, s_i^A)\}_{i=1}^{N_A},
\end{equation}
where $X_i^A \in \mathbb{R}^{T \times F}$ is a Mel spectrogram segment, $y_i^A \in \{0,1\}$ is the OSA label, and $s_i^A \in [0,1]$ is the night-specific SpO\textsubscript{2} time percentage, $N_A$ denotes the number of night recordings. Let the paediatric dataset be 
\begin{equation}
\mathcal{D}_C = \{(X_j^C, y_j^C, s_j^C)\}_{j=1}^{N_C}, \quad N_C \ll N_A .
\end{equation}

We consider a CNN-based model $f_\theta(X) = (\hat{y}, \hat{s})$, where $\hat{y} \in [0,1]$ is the predicted OSA probability and $\hat{s} \in [0,1]$ is the predicted SpO\textsubscript{2} desaturation. The model includes either a single-task head (predicting only OSA) or a multi-task head (jointly predicting OSA and SpO\textsubscript{2}). Delayed SpO\textsubscript{2} labels are defined with a time delay 
\begin{equation}
\Delta t = t_{\text{OOD}} - t_{\text{ONAC}}, 
\end{equation}
where $t_{\text{ONAC}}$ is the onset of nasal airflow cessation and $t_{\text{OOD}}$ is the onset of a 3\% oxygen desaturation.

This work transfers an adult-trained model to the paediatric domain by systematically evaluating the impact of different strategies, including task type, fine-tuning approach, and the use of SpO\textsubscript{2} labels. Research questions are formulated as:\\
\textbf{RQ1:} Does fine-tuning, particularly in multi-task learning, improve paediatric OSA detection compared to single-task learning and models without fine-tuning?\\
\textbf{RQ2:} How do fine-tuning strategies (frozen encoder, full fine-tuning) and the use of SpO2 labels (delayed versus immediate labels) influence model performance in paediatric OSA detection?\\
\textbf{RQ3:} Which combination of task type, fine-tuning strategy and SpO2 labels provides the most reliable and clinically meaningful predictions for paediatric OSA severity?
% Let $M_y(f,\mathcal{D})$ denote the OSA classification performance of model $f$ on dataset $\mathcal{D}$. 
% Model variants are denoted $f_{\theta}^{\tau,g,\ell}$, where 
% $\tau\in\{\text{single},\text{multi}\}$ is the task type, 
% $g\in\{\text{NoFT},\text{FrozenEnc},\text{FullFT}\}$ is the fine-tuning strategy, 
% and $\ell\in\{\text{immediate},\text{delayed}\}$ indicates the SpO$_2$ label usage.  

% \textbf{RQ1.} Does fine-tuning, especially in multi-task learning, improve paediatric OSA detection?
% \begin{equation}
% M_y\!\big(f_{\theta}^{\tau,g,\ell},\mathcal{D}_C\big) \;>\; M_y\!\big(f_{\theta}^{\tau,\text{NoFT},\ell},\mathcal{D}_C\big),
% \end{equation}
% \begin{equation}
% M_y\!\big(f_{\theta}^{\text{multi},g,\ell},\mathcal{D}_C\big) \;>\; M_y\!\big(f_{\theta}^{\text{single},g,\ell},\mathcal{D}_C\big).
% \end{equation}

% \textbf{RQ2.} How do fine-tuning strategies $g$ and SpO$_2$ label types $\ell$ influence performance?
% \begin{equation}
% (g^\ast,\ell^\ast) \;=\; \arg\max_{g,\ell} \; M_y\!\big(f_{\theta}^{\text{multi},g,\ell},\mathcal{D}_C\big).
% \end{equation}

% \textbf{RQ3.} Which overall configuration yields the best paediatric OSA performance?
% \begin{equation}
% (\tau^\ast,g^\ast,\ell^\ast) \;=\; \arg\max_{\tau,g,\ell} \; M_y\!\big(f_{\theta}^{\tau,g,\ell},\mathcal{D}_C\big).
% \end{equation}

\subsection{Physiological information} 
\label{Physiological information}

\begin{figure}[thb]
    \centering
    \includegraphics[width=1\linewidth]{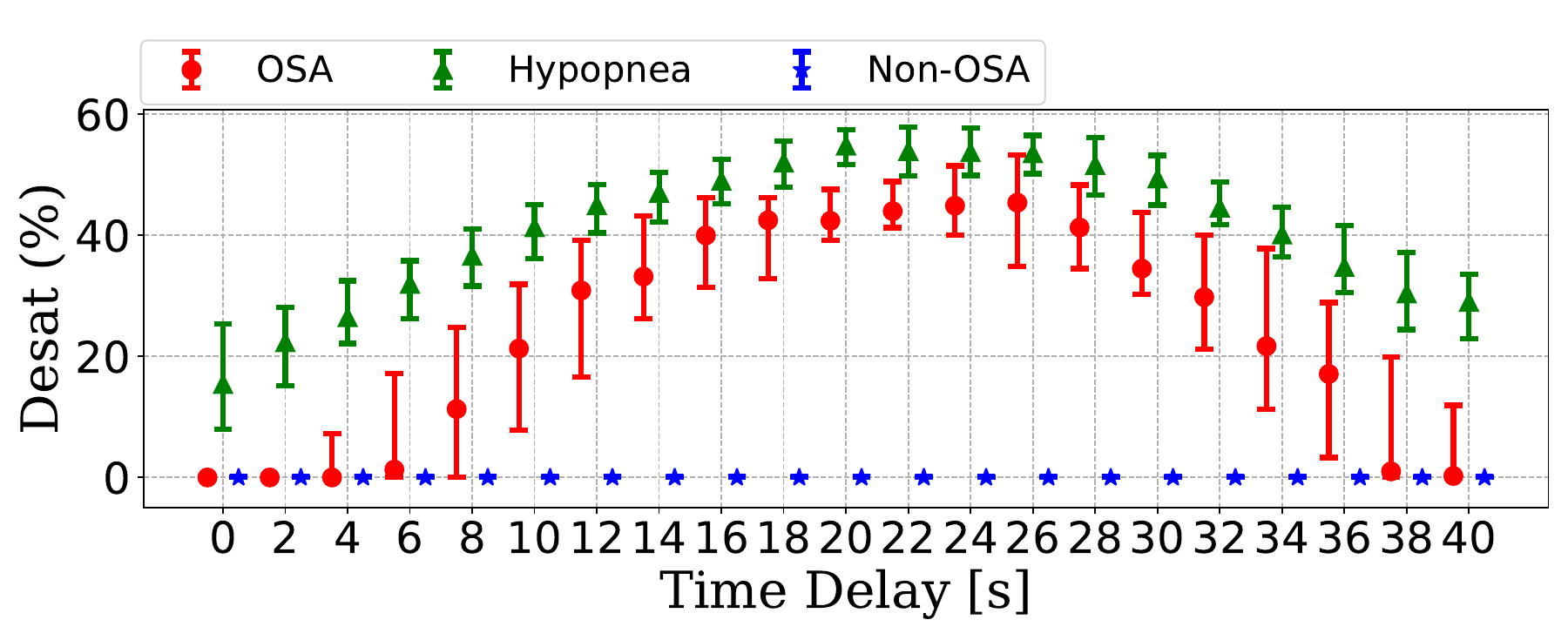}
    \caption{Median - 95\% CI bar plots showing time spent below SpO\textsubscript{2} baseline as a percentage within a 15\,s window versus time delay of SpO\textsubscript{2} for non-OSA, OSA and hypopnoea segments. Median and 95\% CI for non-OSA events are 0.}
    %are plotted, indicating that a 26s delay maximises strong discrimination between OSA and Non-OSA events.}
    \label{fig:Desaturation-vs-Delay}
    % \caption{Time delay between SpO\textsubscript{2} and apnoea events is illustrated (top). The time spent below OOD within a 15s window is shown as a percentage for normal and abnormal sleep across varying time delays between ONAC and OOD (bottom).}
    % \label{fig:combined-delay}
\end{figure}

Analysis of paired acoustic and SpO\textsubscript{2} segments revealed that OSA events show larger desaturation (below 3\% oxygen desaturation) proportions than non-OSA events, and that incorporating a temporal delay enhances discrimination (see Fig.~\ref{fig:Desaturation-vs-Delay}).
% 30\,s mel spectrum segments with corresponding 15\,s desaturation windows (see Fig.~\ref{fig:delay}) showed that OSA events exhibit a greater proportion of time spent below 3\% oxygen desaturation compared to non-OSA events (see Fig.~\ref{fig:Desaturation-vs-Delay} at time delay of 0), and that increasing such time delay further enhances the discrimination between the two (see high delays in Fig.~\ref{fig:Desaturation-vs-Delay}). This finding aligns with prior work \cite{ng2006using}. A median time delay of 26s across all patients was thus incorporated into the multi-task learning pipeline, alongside night-specific delays per participant (median varying from 0 to 36s), to evaluate their impact on model performance.
Consistent with prior physiological findings \cite{ng2006using}, a median time delay of 26\,s was observed across patients (range: 0–36\,s). This delay was embedded into the multi-task learning pipeline, with both global and subject-specific delays evaluated to assess their impact on classification performance.

\subsection{Transfer learning strategy} \label{Transfer learning strategy}

We decompose the adult CNN into an \emph{encoder} $g_\theta$ that extracts feature embeddings from a spectrogram segment, and a \emph{prediction head} $h_\phi$ that maps embeddings to outputs: $g_\theta(X) \in \mathbb{R}^d, \; h_\phi(g_\theta(X)) = (\hat{y}, \hat{s}) \in \mathbb{R}^2$ (see Fig.~\ref{fig:Pipeline}). Two transfer learning strategies are considered in this study:

% In \textit{Strategy 1 (Frozen encoder)}, the encoder $g_\theta$ is fixed and only the head $h_\phi$ is trained on paediatric data, with optimised parameters 
% \begin{equation}
% \phi^* = \arg \min_\phi \tfrac{1}{N_C} \sum_{j=1}^{N_C} [\, \mathcal{L}_{\mathrm{cls}}(y_j^C, \hat{y}_j) + \mathcal{L}_{\mathrm{reg}}(s_j^C, \hat{s}_j) \,],
% \end{equation}
% where $(\hat{y}_j, \hat{s}_j) = h_\phi(g_\theta(X_j^C))$. 

% In \textit{Strategy 2 (Full fine-tuning)}, both encoder and head are updated, with optimised parameters 
% \begin{equation}
% (\theta^*, \phi^*) = \arg \min_{\theta,\phi} \tfrac{1}{N_C} \sum_{j=1}^{N_C} [\, \mathcal{L}_{\mathrm{cls}}(y_j^C, \hat{y}_j) + \mathcal{L}_{\mathrm{reg}}(s_j^C, \hat{s}_j) \,],
% \end{equation}
% where $(\hat{y}_j, \hat{s}_j) = h_\phi(g_\theta(X_j^C))$. $\theta^*$ and $\phi^*$ represent the encoder and head parameters, respectively, after fine-tuning on the children dataset. 

In \textit{Strategy 1 (Frozen encoder)}, the encoder $g_\theta$ is fixed and only the head $h_\phi$ is trained on paediatric data, with optimised parameters  
\begin{equation}
\phi^* = \arg \min_\phi \; \frac{1}{N_C} \sum_{j=1}^{N_C} \Big[ \mathcal{L}_{\text{BCE}}(y_j^C, \hat{y}_j) + \mathcal{L}_{\text{MSE}}(s_j^C, \hat{s}_j) \Big],
\end{equation}
where $(\hat{y}_j, \hat{s}_j) = h_\phi(g_\theta(X_j^C))$ and $\mathcal{L}_{\text{BCE}}$ and $\mathcal{L}_{\text{MSE}}$ denote binary cross-entropy and mean squared error losses, respectively.  

In \textit{Strategy 2 (Full fine-tuning)}, both encoder and head are updated:  
\begin{equation}
(\theta^*, \phi^*) = \arg \min_{\theta,\phi} \; \frac{1}{N_C} \sum_{j=1}^{N_C} \Big[ \mathcal{L}_{\text{BCE}}(y_j^C, \hat{y}_j) + \mathcal{L}_{\text{MSE}}(s_j^C, \hat{s}_j) \Big],
\end{equation}
with $\theta^*$ and $\phi^*$ denoting the encoder and head parameters after fine-tuning on the paediatric dataset.

\begin{table*}[t]
\centering
\footnotesize
\caption{Participant demographics and comparison of model performance (MAE and RMSE) across patients under various conditions. \textit{SNR} reflects signal quality, where higher values indicate cleaner snore signals. \textit{Target AHI} is based on home sleep tests scored by qualified sleep technologists. AHI Prediction: S/M: Single-/Multi-Task Learning; NF/F/UF: No Fine-Tuning/Freezing/Un-Freezing Encoder; FD/SD: Fixed Delay (global median time delay)/Specific Delay (night-specific time delay). Predicted values that are closest to the target value are in bold. }
%MAE and RMSE are computed between the scored AHI and the AHI prediction of model variants.}
\label{Tab:15_results}
\begin{tabular}{cccccc|c|cccccccccc}
\toprule
\multicolumn{6}{c|}{Patient Demographics and Signal Attributes} &  Target& \multicolumn{10}{c}{Predicted AHI} \\

\cmidrule(lr){1-6}
\cmidrule(lr){8-17}

ID & Sex & Age & BMI   & Dur & SNR  & AHI & S- & S-   & S-  & M- & M-  & M-  & M-F & M-UF & M-F & M-UF \\

 &  &   &   & (hrs)& (dB)  &  & NF & F   & UF  & NF & F  & UF  & -FD & -FD & -SD & -SD  \\

\midrule
003     & F   & 5   & 15.1  & 9.3 & 1.2 & 4          & 0.11 & 11.29 & 9.78  & \textbf{1.18} & 0.11 & 0.32  & 0.11   & 0.11    & 0.11   & 0.75    \\

004     & M   & 10  & 31.4  & 8.2  & 7.7  & 3.4        & 0.12 & 9.63  & 14.27 & 6.95 & 0.12 & \textbf{1.83}  & 0.12   & 0.12    & 0.12   & 7.44    \\

005     & M   & 5   & 15.9  & 8.7  & -3.1  & 9.4        & 0.11 & 1.49  & \textbf{5.17}  & 3.79 & 0.34 & 0.57  & 0.11   & 0.11    & 0.11   & 1.03    \\

006     & F   & 2   & 17.8  & 10   & 5.5  & 0.8        & 0.6  & 9.5   & 4.1   & 0.3  & 0.3  & 9.7   & 0.3    & 0.1     & 5.5    & \textbf{0.7}     \\

007     & M   & 11  & 33.1  & 7.8  & 1.5  & 6.4        & 0.13 & 1.15  & 0.64  & 2.18 & 0.64 & 6.41  & 2.31   & 0.13    & \textbf{6.41}   & 0.13    \\

008     & F   & 9   & 22.99 & 7.7    & -0.5 & 1.8        & 0.13 & 0.65  & 0.65  & \textbf{1.43} & 0.13 & 0.65  & 0.13   & 0.13    & 0.13   & 0.13    \\

009     & M   & 1   & 19.44 & 9.8  & -6.8  & 8  & 0.1  & 1.53  & 1.53  & 0.51 & 0.71 & \textbf{2.55}  & 0.1    & 0.1     & 0.1    & 0.1     \\

010    & M   & 1   & 16.77 & 8.8  & -0.4  & 1.6        & 0.23 & 0.11  & 0.34  & 0.11 & \textbf{1.48} & 0.34  & 0.34   & 0.11    & 0.11   & 0.11    \\

011    & F   & 13  & 47.28 & 6.1   & 21  & 1.8        & 0.16 & \textbf{1.8}   & 3.44  & 0.49 & 4.75 & 9.34  & 2.46   & 0.16    & 8.03   & 1.8     \\

012    & M   & 1   & 18.62 & 6.9 & -0.4  & 2.6        & 0.14 & 1.3   & 0.43  & 1.88 & 1.3  & 0.14  & 0.43   & \textbf{3.04}    & 0.72   & 0.14    \\

013    & M   & 5   & 17.08 & 8.3    & 1.4   & 5.3        & 0.12 & 0.36  & 2.53  & 0.12 & 1.33 & 1.57  & 3.25   & 4.46    & 2.53   & \textbf{4.46}    \\

014    & M   & 8   & 28.45 & 9.6  & -4.8   & 21         & 0.1  & 1.56  & 0.31  & 0.1  & 19.9 & 14.48 & \textbf{20.52}  & 30.1    & 18.23  & 23.85   \\

015    & M   & 4   & 18.44 & 9.1   & -18.9   & 0.4        & 0.33 & 1.43  & 0.77  & 0.33 & 0.55 & 0.77  & \textbf{0.33}   & 0.11    & 0.11   & 0.11    \\

016    & F   & 15  & 41.61 & 5   & 13.0 & 0.8        & 0.2  & 1     & 16.2  & 1    & \textbf{0.6}  & 0.2   & 8.2    & 0.2     & 3.8    & 1.4     \\

017    & M   & 6   & 32.15 & 8.6 & -1.3   & 2.1        & 0.12 & 0.35  & 0.35  & \textbf{2.21} & 0.35 & 0.35  & 0.12   & 0.12    & 0.12   & 0.12    \\
\midrule
   &     &     &       &    &      &  MAE   & 4.45 & 4.88  & 5.57  & 3.64 & 2.87 & 3.59  & 3.11   & 3.29    & 3.41   & \textbf{2.81}    \\
  &     &     &       &    &   &  RMSE  & 6.81 & 6.89  & 7.92  & 6.3  & 3.9  & 4.68  & 4.2    & 4.55    & 4.26   & \textbf{3.86}   \\
\bottomrule
\end{tabular}
\end{table*}

\section{EXPERIMENTAL SETUP}
\label{s:expr}

% \subsection{System setup}

%DESCRIBE DIFFERENT SYSTEMS HERE. WHAT HAVE BEEN INCLUDED IN EXPERIMENTS?

We first established two baselines: a single-task learning (STL) model and a multi-task learning (MTL) model trained on the adult dataset without fine-tuning, directly mapping acoustic features to OSA labels. The pretrained model jointly predicts OSA probability and SpO\textsubscript{2} desaturation, with the regression branch omitted in STL (Fig.~\ref{fig:Pipeline}).

Two fine-tuning strategies were then applied to the pretrained model (Section~\ref{Transfer learning strategy}), yielding four variants across STL and MTL formulations. Within the MTL framework, we further incorporated SpO\textsubscript{2} desaturation time-delay information during fine-tuning (Section~\ref{Physiological information}). Two settings were considered: (1) a global median delay computed across all 15 paediatric subjects, and (2) subject-specific median delays computed per night. These produced four additional variants. In total, ten model configurations were evaluated.

% We then considered two fine-tuning strategies (see Section \ref{Transfer learning strategy}) applied to the pretrained model. Two strategies yield four variants across STL and MTL models, enabling comparison of fine-tuning strategies and task formulation (STL versus MTL).

%Within the MTL framework, we further incorporated SpO$_2$ time delay information during fine-tuning. Two time delay settings were examined (see Section \ref{Physiological information}): (1) a global median time delay aggregated across all 15 paediatric subjects, and (2) subject-specific median delays computed per night, yielding one specific delay per patient. These settings produce four additional model variants, which allow us to assess the effect of SpO$_2$ delay modelling on performance. In total, we trained ten model variants, from which the best-performing configuration is selected.

\subsection{Training setup}

% I've moved the sentence below to the previous section.
%The pretrained model predicts OSA probability and SpO\textsubscript{2} desaturation, trained on the adult dataset; single-task models omit the regression term (see Fig.~\ref{fig:Pipeline}). 

Adult models were trained for 50 epochs with a batch size of 1024 and a learning rate of $1 \times 10^{-3}$ using the Adam optimiser. For paediatric adaptation, 15 nights of children’s Mel spectrograms with dual labels were fine-tuned for 20 epochs (batch size 8, learning rate $1 \times 10^{-5}$, early stopping 5 epochs), balancing gradual updates and gradient noise to avoid overfitting. Models were implemented in TensorFlow Keras and trained on an RTX 8000 GPU (48GB).

\subsection{Evaluation framework}

% The standard clinical measure for assessing OSA is the apnoea-hypopnoea index (AHI). This is defined as the number of apnea and hypopnea events per hour of sleep: 
% \begin{equation}
% \text{AHI} = \frac{N_{\text{apnea}} + N_{\text{hypopnea}}}{T_{\text{sleep}}}
% \end{equation}
% where $N_{\text{apnea}}$ and $N_{\text{hypopnea}}$ denote the number of apnoea and hypopnoea events, and $T_{\text{sleep}}$ is the total sleep time in hours. 
Predicted AHI was derived by aggregating the model-predicted OSA events across all segments and normalising by total sleep time. 
%In paediatric patients, OSA severity is classified based on AHI thresholds: $<$5 as normal/mild, 5–10 as moderate, and $>$10 as severe \cite{gourishetti2021baseline}. 
We used 5-fold cross-validation on the 15-night paediatric dataset, using 12 nights for fine-tuning (9 were used for fine-tuning and 3 were used for validation) and 3 for testing in each fold. This ensures that training and test data come from different patients, providing cross-patient validation. Model performance was assessed using mean absolute error (MAE) and root mean square error (RMSE) between the predicted AHI and the reference AHI from the scored HSAT data.

\section{RESULTS AND DISCUSSION}
\label{s:results}

Table~\ref{Tab:15_results} summarises AHI prediction performance across model variants. Fine-tuning consistently improved performance compared with models trained only on adult data across the whole data set in terms of MAE and RMSE measures. Without adaptation, the adult STL model produced near-zero predictions for most patients, effectively classifying all cases as normal and failing to reflect OSA severity. In contrast, fine-tuned models generated higher and more variable AHI estimates that better matched the reference values. Across all configurations, multi-task learning (MTL) generally outperformed single-task learning (STL), indicating that jointly modelling OSA classification and SpO\textsubscript{2} regression provides more informative representations.

Incorporating temporally delayed SpO\textsubscript{2} labels further reduced prediction error, with night-specific delays yielding the largest improvements. Among models that did not use night-specific delays, freezing the encoder performed comparably to, or slightly better than, full fine-tuning, suggesting that preserving pretrained acoustic representations is beneficial when physiological alignment is less precise. The overall best performance was achieved by combining MTL, full fine-tuning, and subject-specific delays (M--UF--SD), which produced the lowest MAE (2.81) and RMSE (3.86).

%\textbf{Clinical relevance}: 
These reductions in AHI prediction error are clinically meaningful, as an AHI $\geq$ 10 typically indicates severe paediatric OSA requiring further diagnostic evaluation. For example, patient ID 014 (reference AHI = 21) was correctly identified as severe by the fine-tuned MTL models, whereas STL and non-adapted models misclassified this case as normal or mild. The M--UF--SD configuration also showed improved severity stratification in lower-AHI cases, producing predictions within the normal/mild range (0.11--1.8) for patients with reference AHI between 0.4 and 4, thereby reducing potential false positives. Moderate cases were generally estimated with smaller error (e.g., ID 013: reference AHI 5.3 vs. predicted 4.46), although some underestimation remained (e.g., IDs 005 and 007). Overall, the results indicate that physiology-informed transfer learning improves the reliability of AHI estimation and may support triage decisions in paediatric OSA screening.

% Limitations
A key limitation of this study is the small paediatric sample size (15 patients), which limits statistical power and may affect generalisability. Variability in age, BMI, and recording quality further increases inter-subject variability. Nevertheless, this constraint reflects a common challenge in paediatric sleep research and directly motivates the proposed approach. By leveraging larger adult datasets for pretraining and incorporating physiological alignment during adaptation, the framework provides a practical strategy for mitigating data scarcity in paediatric OSA modelling.

\section{Conclusion}
\label{s:conc}

This paper presented a physiology-informed transfer learning framework for estimating paediatric OSA severity from smartphone-recorded breathing sounds. Knowledge learned from adult acoustic data was adapted to paediatric recordings through fine-tuning, with oxygen desaturation incorporated as an auxiliary physiological task. We systematically compared single- and multi-task learning formulations, evaluated encoder freezing versus full fine-tuning, and examined the effect of modelling temporal delays between acoustic events and SpO\textsubscript{2} desaturation. Results showed that multi-task fine-tuning with subject-specific delay modelling achieved the most accurate AHI estimation, demonstrating the value of integrating physiological priors into transfer learning. These findings support the feasibility of non-invasive, child-friendly, smartphone-based screening as a complementary tool for paediatric OSA assessment.

Although the limited size of the paediatric dataset remains a constraint, the proposed framework illustrates how transfer learning can enable data-efficient development of clinically relevant models. Future work will incorporate additional low-cost physiological signals, such as accelerometry and respiratory effort (the thorax and IMU motion sensors illustrated in Fig.~\ref{fig:Pipeline}), and validate the approach in larger and more diverse paediatric cohorts to further support clinical translation.

%\vfill\pagebreak

\bibliographystyle{IEEEbib}
\bibliography{ref}

\end{document}